\documentclass[preprint2]{aastex}
\shorttitle{Is there evidence for dark energy evolution?}
\shortauthors{Ding et al.}

\begin{document}

%% LaTeX will automatically break titles if they run longer than
%% one line. However, you may use \\ to force a line break if
%% you desire.

\title{Is there evidence for dark energy evolution?}

%% Use \author, \affil, and the \and command to format
%% author and affiliation information.
%% Note that \email has replaced the old \authoremail command
%% from AASTeX v4.0. You can use \email to mark an email address
%% anywhere in the paper, not just in the front matter.
%% As in the title, use \\ to force line breaks.
\author{Xuheng Ding$^{1}$, Marek Biesiada$^{1,2}$, Shuo Cao$^{1}$, Zhengxiang Li$^{1}$, and Zong-Hong Zhu$^{1}$}
\affil
{
  $^1$ Department of Astronomy, Beijing Normal University,
    Beijing 100875, China
}
\affil
{
  $^2$ Department of Astrophysics and Cosmology, Institute of Physics, University of Silesia, Uniwersytecka 4, 40-007, Katowice, Poland
}

%% Notice that each of these authors has alternate affiliations, which
%% are identified by the \altaffilmark after each name.  Specify alternate
%% affiliation information with \altaffiltext, with one command per each
%% affiliation.

%% Mark off your abstract in the ``abstract'' environment. In the manuscript
%% style, abstract will output a Received/Accepted line after the
%% title and affiliation information. No date will appear since the author
%% does not have this information. The dates will be filled in by the
%% editorial office after submission.

\begin{abstract}
Recently, \citet{Sahni2014} combined two independent measurements of $H(z)$ from BAO data with the value of the Hubble constant $H_0 = H(z=0)$, in order to test the cosmological constant hypothesis by means of an improved version of the $Om$ diagnostic.
Their result indicated a considerable tension between observations and predictions of the
$\Lambda$CDM model. However, such strong conclusion was based only on three measurements of $H(z)$.
This motivated us to repeat similar work on a larger sample.
By using a comprehensive data set of 29 $H(z)$, we find that
discrepancy indeed exists. Even though the value of $\Omega_{m,0} h^2$ inferred from $Omh^2$ diagnostic depends on the way
one chooses to make a summary statistics (weighted mean or the median), the persisting discrepancy supports the claims
of \citet{Sahni2014} that $\Lambda$CDM model may not be the best description of our Universe.

\end{abstract}


\section{Introduction}
The discovery of accelerating expansion of the Universe \citep{Riess98, Perlmutter99} brought us a mystery
which became one of the most important challenges for modern cosmology and theoretical physics.
This phenomenon has been confirmed since then in manifold ways using different probes like supernovae Ia, acoustic
peaks in the CMBR ~\citep{CMB1, CMB2}, baryon acoustic oscillations (BAO) ~\citep{BAO, Eisenstein11} in the distribution of the large scale structure etc.
By now all this observational evidence was concordant with the simplest assumption that there exists non-vanishing
cosmological constant $\Lambda$. While being the simplest, such a model is not theoretically satisfactory. There is a huge
discrepancy if one tries to motivate $\Lambda$ as a zero-point quantum vacuum energy. Therefore alternative explanations appeared
invoking the scalar field which settled down in an attractor \citep{Ratra88}. This motivated to push forward the phenomenological
picture of the so called dark energy described as a fluid with barotropic equation of state $p = w \rho$, where $w$ can either
be constant -- the so called ``quintessence''\citep{quintessence}(as a value characteristic for the fixed point attractor) or evolving in time \citep{Chevalier01,Linder03} since the scalar field could be expected to evolve in time.
The main drawback of such an approach is that it makes explicit assumption about the dark energy before its specific model (a quintessence or evolving equation of state) can be tested on observational data. Moreover, alternatives to dark energy such like modified gravity~\citep{DGP, FR1, FR2, FT} cannot be easily tested within this phenomenology. All observational tests of quintessence pin-point its value close to $w=-1$ (within the uncertainties) which is equivalent to the cosmological constant. On the other hand tests of varying in time cosmic equation of state are much less restrictive and do not allow to make a decisive statement whether dark energy equation of state evolved or not.

Therefore, we clearly need alternative probes: capable to discriminate between the cosmological constant and evolving dark energy not relying on the dark energy assumption and its equation of state parametrization. One promising approach to such probes has been initiated by \citet{Sahni2008} and developed further in \citep{Sahni2012}. By properly rearranging the equation for the Hubble function in the flat $\Lambda$CDM model: $H^2(z) = H_0^2 [ \Omega_m (1+z)^3 + 1 - \Omega_m ] $, they noticed that the so called $Om(z)$ diagnostic (where ${\tilde h} = H(z) / H_0$):
\begin{equation} \label{Omz}
Om(z) = \frac{{\tilde h}^2(z)-1}{(1+z)^3 - 1}
\end{equation}
 should be constant and equal exactly to the present mass density parameter, if the $\Lambda$CDM model is the true one: $Om(z)_{\Lambda CDM} = \Omega_{m,0}$. This is remarkable and differentiates between $\Lambda$CDM and other dark energy models (including evolving dark energy).
Let us remark that essentially the same idea has also been formulated by \citet{Zunckel2008} who called it ``a litmus test'' for the canonical $\Lambda$CDM model.
Developing this method \citep{Sahni2012} they also considered a generalized two-point diagnostics
\begin{equation} \label{Omz2}
Om(z_1,z_2) = \frac{{\tilde h}^2(z_1)-{\tilde h}^2(z_2)}{(1+z_1)^3 - (1+z_2)^3}
\end{equation}
which should also be equal to $\Omega_{m,0}$ within the $\Lambda$CDM model but has an advantage that having $H(z)$ measurements at $n$ different redshifts, one has $\frac{n(n-1)}{2}$ two point diagnostics, hence a considerably increased sample for inference.

In their latest paper \citet{Sahni2014} used three accurately measured values of $H(z)$ to perform this test. These were: the $H(z=0)$ measurement by \citep{Riess2011,PlanckXVI}, $H(z=0.57)$ measurement from SDDS DR9 \citep{Samushia} and the latest $H(z)=2.34$ measurement from the $Ly{\alpha}$ forest in SDSS DR11 \citep{Delubac}. They found that all three values of the two-point diagnostics $Om(z_1,z_2)h^2$ were in strong tension with the $\Omega_{m,0}h^2$ reported by Planck \citep{PlanckXVI}.  It has been noticed \citep{Delubac, Sahni2014} that such result could be in tension not only with the $\Lambda$CDM model but with other dark energy models based on the General Relativity. Because this conclusion could be of a paramount importance for the dark energy studies, an update of this test with a larger sample of $H(z)$ is essential.

\section{Data, Methodology and Results} \label{sec:method}

\begin{figure*}
\begin{center}
\includegraphics[angle=0,width=12cm]{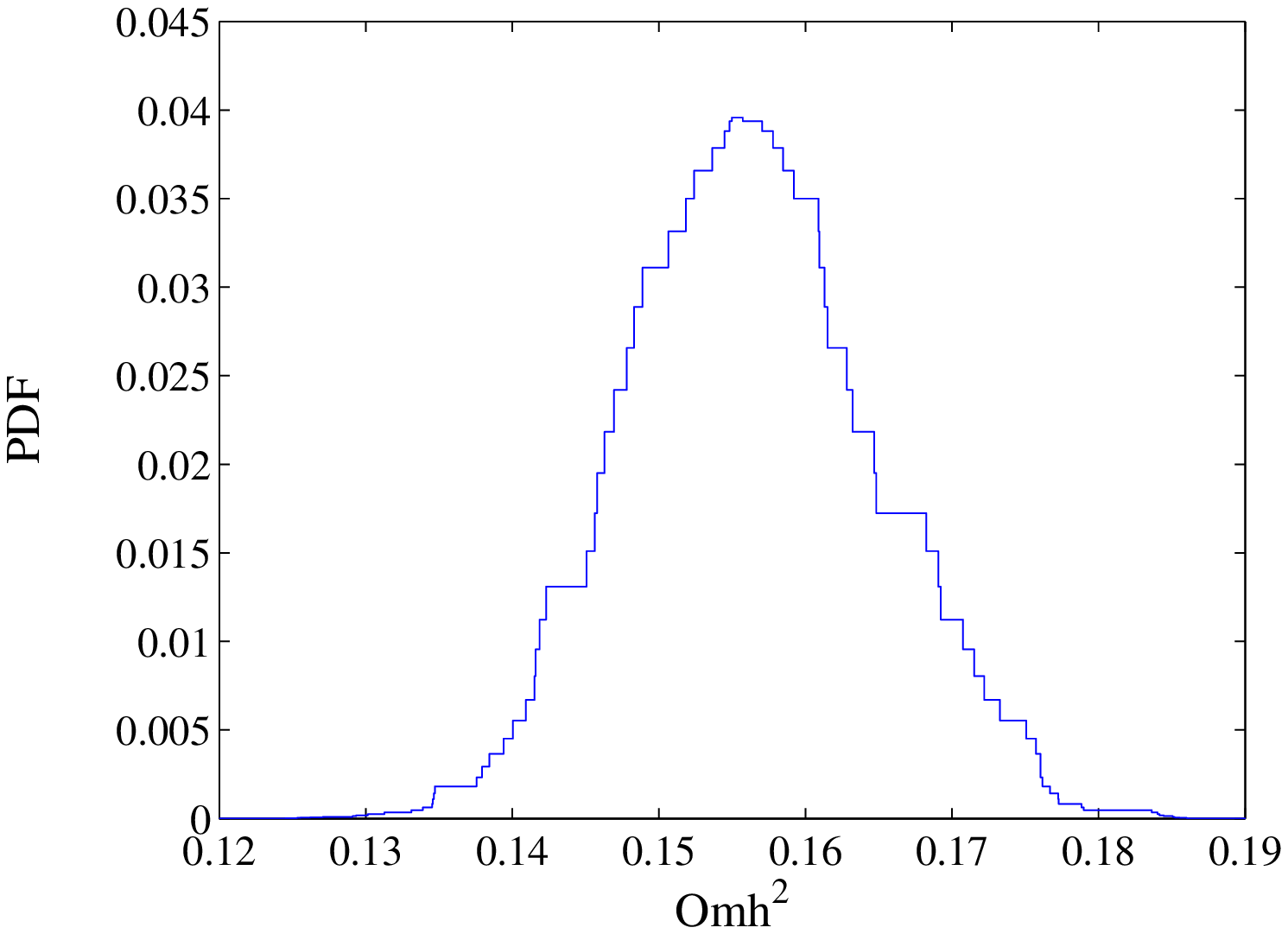}
\caption{Histogram of $Omh^2$ calculated from pairwise $Om(z_1,z_2)h^2$ statistics performed on a sample of 29 measurements of $H(z)$ based on differential ages of galaxies and BAO.}

 \label{fig1}
\end{center}
\end{figure*}

As a basic dataset we used a sample of 29 $H(z)$ measurements taken from the compilation of \citet{Chen2014} modified in the following way: one data point at $z=0.6$ coming from \citet{Blake2012} was added and two data points from \citet{Gatzanaga} were withdrawn. The reason for deleting aforementioned two points is that these results have been debated in subsequent papers e.g. by \citet{Miralda-Escude}, \citet{Kazin} or \citet{Cabre}. For the sake of consistency with \citet{Sahni2014}, we have also taken the latest BAO measurement by \citet{Delubac} $H(z=2.34) = 222\pm 7$ instead of $H(z=2.3) = 224\pm 8$ of \citet{Busca2012}. After these changes our data are essentially like the ones used by \citet{Farooq} with \citet{Busca2012} measurement replaced by \citet{Delubac}.
During the analysis we also made assessments on subsamples of this biggest one, as we will explain further. Part of the data comes from cosmic chronometers -- spectroscopy of galaxies assumed to evolve passively \citep{JimenezLoeb}. Hereafter, this differential age
approach will be quoted as DA for short. The other part of the data comes from BAO -- including the data points used by \citet{Sahni2014}.  Data are summarized in Table~1. Then we proceed in exactly the same way as \citet{Sahni2014}, i.e. for each pair of redshifts $(z_i,z_j)$ we calculate the improved $Om$ diagnostic:
\begin{equation} \label{improved}
Omh^2(z_i,z_j) = \frac{h^2(z_i)- h^2(z_j)}{(1+z_i)^3 - (1+z_j)^3}
\end{equation}
where: $h(z) = H(z) / 100km/sec/Mpc$ is dimensionless Hubble parameter. In particular case of $\Lambda$CDM model the improved diagnostic Eq.~(\ref{improved}) should be equal to $\Omega_{m,0}h^2$ which luckily is the quantity best constrained by the CMBR data, e.g. from Planck \citep{PlanckXVI}. Because the sample of $29$ data points leads to $406$ different pairs, we summarize our calculations on Figure~\ref{fig1}. One can see that the distribution of inferred $\Omega_{m,0}h^2$ is skewed and centered around the value different from this reported by \citet{Sahni2014}.
%%%%%%%%%%%%
\begin{deluxetable}{llll} \label{table1}
\tablewidth{0pt} \tablecaption{Data of the Hubble parameter $H(z)$ versus the redshift $z$, where $H(z)$ and $\sigma_{H}$ are in km s$^{-1}$ Mpc$^{-1}$. These are essentially the data of \citet{Farooq}, with the BAO measurement at the largest redshift $H(z=2.34)$ taken after \citet{Delubac}.
}
\tablehead{ \colhead{$z$} & \colhead{$H(z)$} & \colhead{$\sigma_{H}$} & \colhead{Method} }

\startdata
0.07 & 69 & 19.6 & DA\\
0.1 & 69 & 12 & DA\\
0.12 & 68.6 & 26.2 & DA\\
0.17 & 83 & 8 & DA\\
0.179 & 75 & 4 & DA\\
0.199& 75 & 5 & DA\\
0.2& 72.9 & 29.6 & DA\\
0.27 & 77 & 14 & DA\\
0.28 & 88.8 & 36.6 & DA\\
0.35 & 82.7 & 8.4 & BAO\\
0.352 & 83 & 14 & DA\\
0.4 & 95 & 17 & DA\\
0.44 & 82.6 & 7.8 & BAO\\
0.48 & 97 & 62 & DA\\
0.57 & 92.9 & 7.8 & BAO\\
0.593& 104 & 13 & DA\\
0.6 & 87.9 & 6.1 & BAO\\
0.68 & 92 & 8 & DA\\
0.73 & 97.3 & 7 & BAO\\
0.781 & 105 & 12 & DA\\
0.875 & 125 & 17 & DA\\
0.88 & 90 & 40 & DA\\
0.9 & 117 & 23 & DA\\
1.037 & 154 & 20 & DA\\
1.3 & 168 & 17 & DA\\
1.43 & 177 & 18 & DA\\
1.53 & 140 & 14 & DA\\
1.75 & 202 & 40 & DA\\
%0.43 & 86.45 & 3.68 & BAO\\ % Gatzanaga
%0.24 & 79.69 & 2.65 & BAO\\ % Gatzanaga
%2.3  & 224 & 8 & BAO\\
2.34  & 222 & 7 & BAO\\
\hline
\enddata
\end{deluxetable}

If one is to make a summary statistics one can do it in two ways. First, straightforward way would be to calculate the weighted mean:
\begin{equation} \label{weighted}
Omh^2_{(w.m.)} = \frac{\sum_{i=1}^{29} \sum_{j \textgreater i} Omh^2(z_i,z_j)/ \sigma^2_{Omh^2,ij}}{\sum_{i=1}^{29} \sum_{j \textgreater i} 1 / \sigma^2_{Omh^2,ij}}
\end{equation}
and the standard deviation:
\begin{equation} \label{SD}
\sigma_{(w.m.)} = \left( \sum_{i=1}^{29} \sum_{j \textgreater i} 1 / \sigma^2_{Omh^2,ij} \right)^{-1/2}
\end{equation}
where:
\begin{equation}
\sigma^2_{Omh^2,ij} = \frac{4(h^2(z_i) \sigma^2_{h(z_i)} + h^2(z_j) \sigma^2_{h(z_j)}) }{((1+z_i)^3 - (1+z_j)^3)^2}
\end{equation}
and $\sigma_{h(z_i)}$ denotes the uncertainty of the $i-th$ measurement. We also assume that redshifts are measured accurately.
This well known and often used approach relies, however, on several strong assumptions: statistical independence of the data, no systematic effects, Gaussian distribution of the errors. These assumptions, especially the Gaussianity of errors are not valid here. Hence the weighted mean of
\begin{equation} \label{weighted mean}
Omh^2_{(w.m.)} = 0.1253 \pm 0.0021
\end{equation}
is not a reliable measure, as one can see form the histogram on Figure~\ref{fig1}. We will comment more on this non-Gaussianity in a moment.

Therefore we took another, much more robust approach: to calculate the median. This approach was pioneered by the paper of \citet{GottIII} and then used by the others, e.g. quite recently by \citep{RatraMedian}. The robustness of this method stems from the fact that if systematic effects are absent, half of the data is expected to be higher and another half - lower than the median. Then, as the number of measurements $N$ increases, calculated median approaches its true value. So the median has clear and robust meaning without need to assume anything about the error distribution. From the definition of the median, probability that any particular measurement, one of $N$ independent measurements is higher than the true median is $50 \%$.
Consequently, the probability that $n$ observations out of the total number of $N$ is higher than the median follows the binomial distribution:
$P = 2^{-N} N! / [n!(N-n)!]$. This allows to calculate in a simple manner the confidence regions (e.g. $68 \%$ confidence region) of the median value estimated from the sample. Proceeding this way we have obtained:
\begin{equation} \label{Median}
Omh^2_{(median)} = 0.1550^{+0.0065}_{- 0.0072}
\end{equation}

\begin{figure*}
\begin{center}
\includegraphics[angle=0,width=70mm]{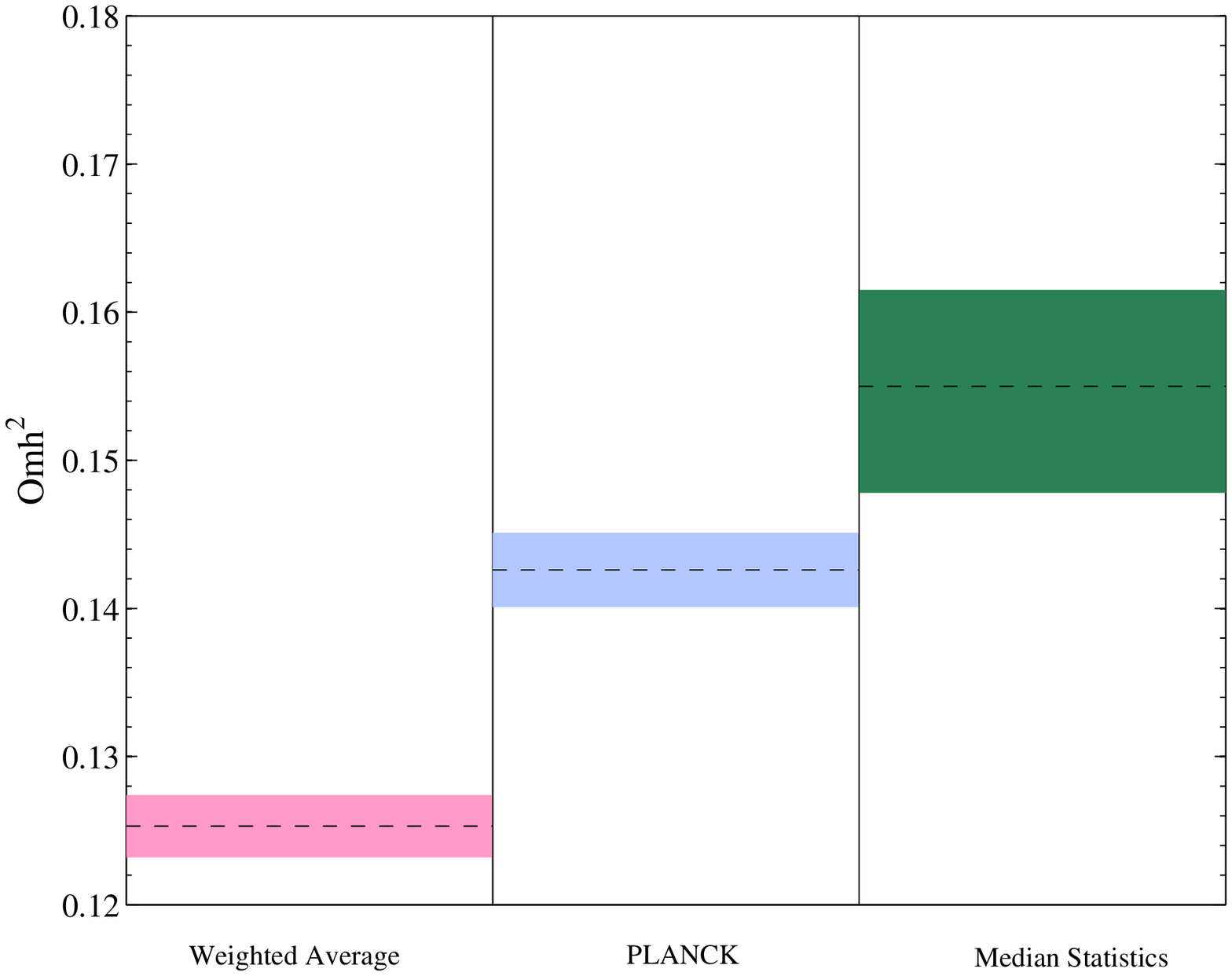}
\includegraphics[angle=0,width=70mm]{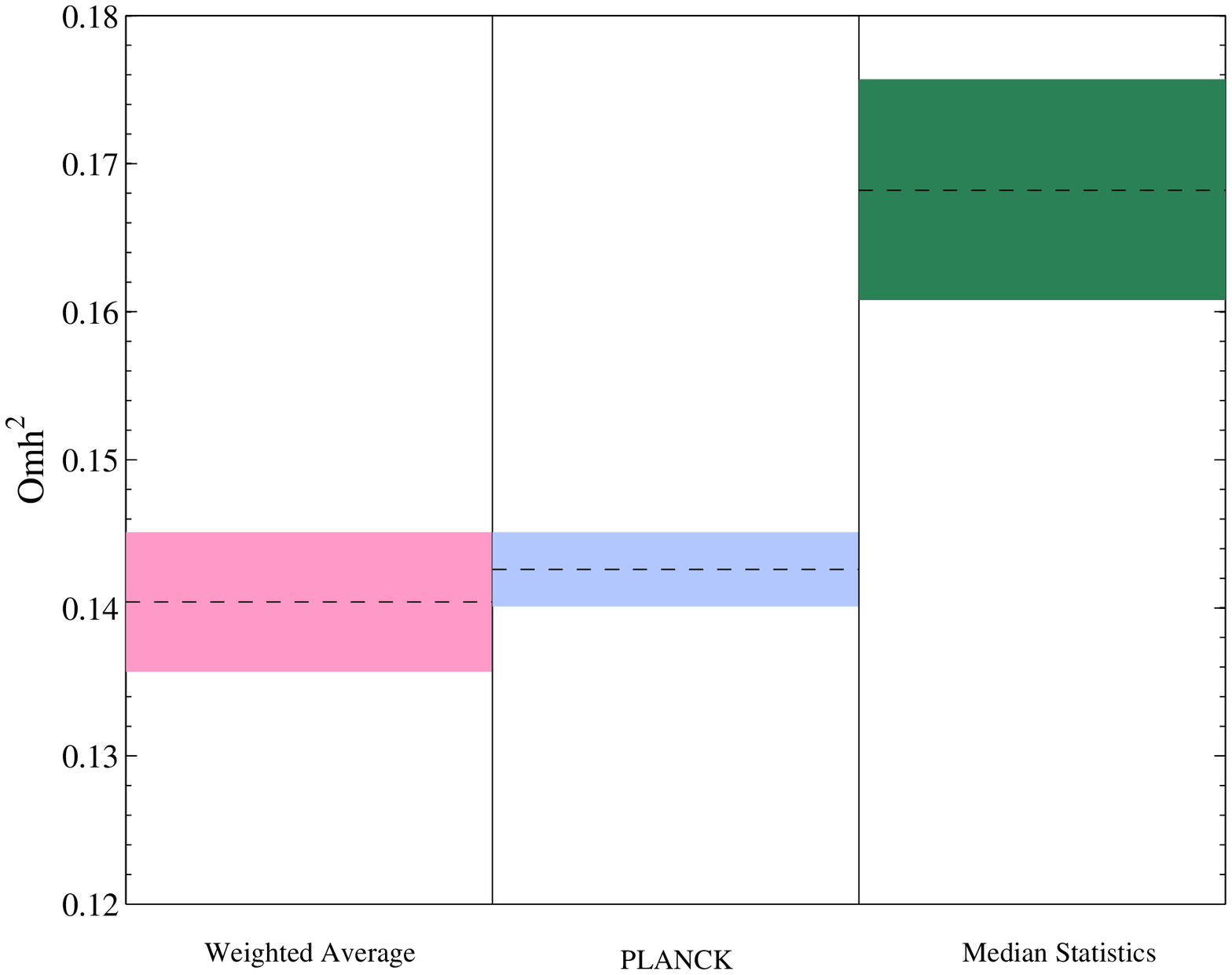}
\includegraphics[angle=0,width=70mm]{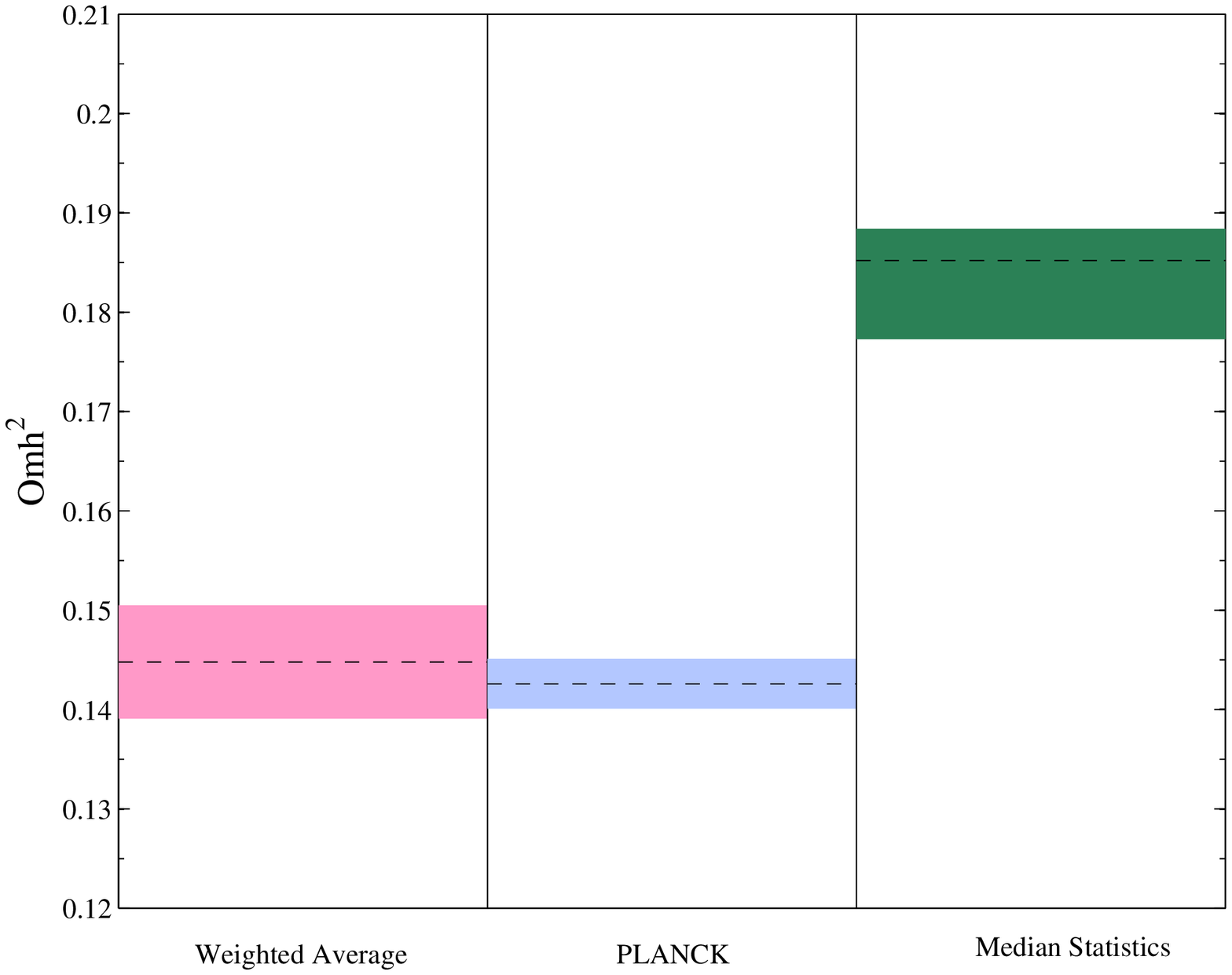}
\includegraphics[angle=0,width=70mm]{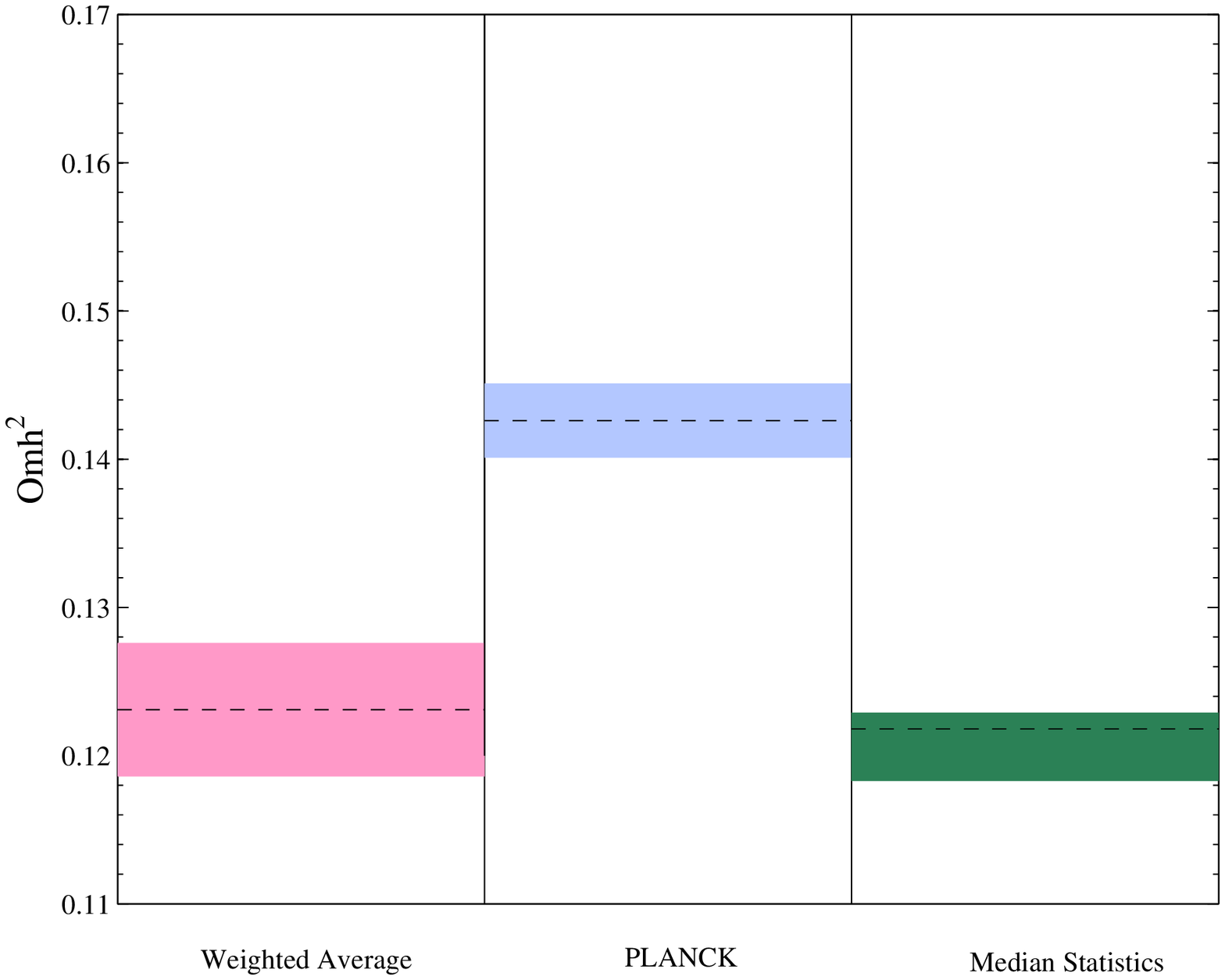}
\caption{Values of $Omh^2$ calculated from the $Omh^2$ diagnostic as a weighted mean of data and as the median value.
The result from Planck experiment is shown for comparison. Bands display the $68 \%$ confidence regions. Upper - left figure corresponds to the full sample $n=29$ data points, upper - right shows the results when $z=2.34$ point was dropped ($n=28$), lower - left corresponds to DA data only ($n=23$) and lower - right figure --- BAO only ($n=6$).}

\label{fig2}
\end{center}
\end{figure*}

In order to facilitate comparison between the inferred values of $\Omega_{m,0}h^2$ obtained from two statistical approaches and the Planck data, we display the results in Figure~\ref{fig2}.

Inconsistency between $Omh^2(z_1,z_2)$ diagnostic calculated on $H(z)$ data and the Planck value of $\Omega_{m,0}h^2$ as well as mutual inconsistency between weighted averaging and median statistics schemes, motivated us to make some more detailed tests. First, we recalculated $Omh^2$ for three sub -- samples: excluding the highest redshift $z=2.34$ measurement, using only DA data and using only BAO data. Results are visualized on Figure~\ref{fig2} and shown in more details in Table~2. One can see that $z=2.34$ point had a big leverage on the weighted average value --- dropping this point one achieves agreement with the $\Lambda$CDM Planck value. However, the question remains whether the weighted average scheme is appropriate.
Therefore, following \citet{Chen} and \citet{Crandall}, we have drawn histograms of distribution of our measurements as a function of the number of standard deviations $N_{\sigma}$ away from central estimates (weighted mean and the median respectively). Because of limited space we do not show them here, but report in Table~2 corresponding percentage of the distribution falling within $\pm 1\sigma$ i.e. $|N_{\sigma}|<1$. One clearly sees that they strongly deviate from the Gaussian $68 \;\%$ expectation. We also tested the $N_{\sigma}$ distribution with Kolmogorov-Smirnov test which strongly rejected the hypothesis of Gaussianity in each sub-sample (with p-values ranging from $10^{-4}$ to $10^{-7}$). Therefore our conclusion is that weighted average scheme is not appropriate here and the median statistics is more reliable. Both statistical methods: weighted mean and median produce similar results for the BAO data, but with addition of DA data these two schemes give drastically different results. This may suggest the existence of some systematic error in DA data. It is not obvious by itself, because the nonlinear relation between input variables ($H(z)$) underlying the $Omh^2$ diagnostic may be the source of asymmetric uncertainties of the latter. However, the fact that BAO and DA data deviate from the $\Lambda$CDM expected result in opposite directions strongly supports suggestion of unaccounted systematics. This will be the subject of a separate study.

\begin{deluxetable}{lllll} \label{table2}
\tablewidth{0pt} \tablecaption{Values of $Omh^2$ diagnostic central values (weighted mean and median) and their ``non-Gaussianity'' indicated by the percentage of distribution falling within $|N_{\sigma}|<1$ for the main sample and different sub-samples. Observations of the CMB from PLACNK inform us that $\Omega_{m,0}h^2_{Planck} = 0.1426 \pm 0.0025.$
}
\tablehead{ \colhead{Sample} & \colhead{$Omh^2_{(w.m.)}$} & \colhead{$|N_{\sigma}(w.m.)|<1$ } & \colhead{$Omh^2_{(median)}$} & \colhead{$|N_{\sigma}(median)|<1$}     }

\startdata
Full sample ($n=29$) & $0.1253 \pm 0.0021$ & $80.54 \%$ & $0.1550^{+0.0065}_{-0.0072}$ & $75.62 \%$   \\
$z=2.34$ excluded ($n=28$)  & $0.1404 \pm 0.0047$ & $77.78 \%$ & $0.1682^{+0.0075}_{-0.0074}$ & $82.80 \%$    \\
DA only ($n=23$) & $0.1448 \pm 0.0057$ & $77.47 \%$ & $0.1852^{+0.0032}_{-0.0079}$ & $86.56 \%$    \\
BAO only ($n=6$)  & $0.1231 \pm 0.0045$ & $100 \%$ & $0.1218^{+0.0011}_{-0.0035}$ & $100 \%$    \\
\hline
\enddata
\end{deluxetable}

\section{Conclusions} \label{sec:conclusions}

In this paper we attempted to assess the $Omh^2(z_1,z_2)$ diagnostic introduced and developed by \citet{Sahni2012}.
The main reason to do so was the recent paper by \citet{Sahni2014} where they claimed that recent precise measurements of expansion rates
at different redshifts suggest a severe tension with the $\Lambda$CDM model.
We repeated this on a much more comprehensive data set of $29$ $H(z)$ obtained by two techniques: DA and BAO.
One can see from the Table~1 that uncertainties of $H(z)$ obtained by different methods are different. Even within the same methodology (DA) uncertainties are different from case to case. The $Omh^2(z_1,z_2)$ diagnostic, involving ratio of certain differences (see Eq.~(\ref{improved})) calculated on our data produces an asymmetric distribution. This means that the weighted mean is not a reliable summary measure. Therefore we used a more robust approach to calculate the median.

Our result is that the value
of $Omh^2$ inferred from $Omh^2$ diagnostic is indeed in tension with the one obtained by Planck (under assumption of the $\Lambda$CDM model).
In our case, this tension is not so severe as in \citet{Sahni2014} ($Omh^2 = 0.122 \pm 0.01$ vs. $\Omega_{m,0}h^2_{Planck} = 0.1426 \pm 0.0025$).
Even though the inferred value is sensitive to the way one chooses to make a summary statistics: the weighted mean value is lower and the median value is higher than that obtained by Planck, they are both discrepant with each other. Non-Gaussianity in the data suggests that median statistics approach is more appropriate, so this tension cannot be alleviated by excluding high redshift data.

This supports the claims of \citet{Sahni2014} that the concordance model ($\Lambda$CDM) might not be the true or even the best one describing our Universe.
Therefore, we also performed a quick test whether XCDM or CPL models (simplest evolving equation of state parametrization) best fitted to the Planck or WMAP9 data agree better with $H(z)$ data. In such a case the $Omh^2(z_i,z_j)$ diagnostic defined be Eq.(\ref{improved}) will no longer be a single number $\Omega_{m,0}h^2$, but rather: $Omh^2(z_i,z_j) = \Omega_{m,0}h^2 + (1 - \Omega_{m,0})h^2 \frac{(1+z_i)^{3(1+w)} - (1+z_j)^{3(1+w)}}{(1+z_i)^3 - (1+z_j)^3}$ for XCDM and
$Omh^2(z_i,z_j) = \Omega_{m,0}h^2 + (1 - \Omega_{m,0})h^2\\ \frac{(1+z_i)^{3(1+w_0+w_a)}e^{\frac{- 3 w_a z_i}{1+z_i}} - (1+z_j)^{3(1+w_0+w_a)}e^{\frac{- 3 w_a z_j}{1+z_j}}}{(1+z_i)^3 - (1+z_j)^3}$
for CPL parametrization. Therefore, for each pair we formed the residuals $R(z_i,z_j)$ by subtracting the right-hand sides from the left-hand sides. Because such residuals inherit non-Gaussianity from $Omh^2(z_i,z_j)$ we summarized our findings with weighted average $R_{(w.m.)}$ and the median  $R_{(median)}$.
If a particular model (XCDM or CPL) agreed better with the $H(z)$ data than $\Lambda$CDM, then its $R$ should have been closer to zero than $R(\Lambda CDM)_{(w.m.)} = -0.0173 \pm 0.0033$ or $R(\Lambda CDM)_{(median)} = 0.0124^{+0.0070}_{-0.0076}$. In the XCDM model we took the $w = -1.0507^{+0.0469}_{-0.0507}$ parameter according to \citet{Cai} best fit to Plack+WMAP9 data.
% their best fit comprise also $\Omega_{m,0} = 0.2936^{+0.0103}_{-0.0110}$ and $h=0.7048^{+0.0123}_{-0.0104}$
For the CPL parametrization of the equation of state, we used the values: $w_0 = -1.17^{+0.13}_{-0.12}$, $w_a = 0.35^{+0.50}_{-0.49}$ best fitted to WMAP+CMB+BAO+$H_0$+SNe according to \citet{WMAP9}.

The result is: $R(XCDM)_{(w.m.)} = -0.0176 \pm 0.0025$ and $R(XCDM)_{(median)} = 0.0151^{+0.0063}_{-0.0068}$. For the CPL varying equation of state parametrization we get
$R(CPL)_{(w.m.)} = -0.0275 \pm 0.0073$ or $R(CPL)_{(median)} = - 0.0517^{+0.0149}_{-0.0077}$. We see that they do not reconcile the discrepancy but their performance is even worse than $\Lambda$CDM. However, it is not a decisive conclusion, because what remains to be done is find the values of equation of state parameters $w$ or $(w_0,w_a)$ best fitted to the $H(z)$ data (according to $Omh^2$ diagnostics). This will be a subject of a separate study.

Having confirmed the discrepancy between $\Lambda$CDM and $H(z)$ data its origin should be studied in greater details. One reason could be that our phenomenological description of accelerated expansion with $\Lambda$CDM is incorrect. On the other hand we pointed out that the conclusion (more specifically the direction of this discrepancy) depends on the statistical approach taken. So one should investigate possible systematics in both methods --- DA and BAO and their effect on the conclusion. This is a subject of an ongoing study.

\section*{Acknowledgements}
The authors are grateful to the referees for very useful comments which allowed to improve the paper.
This work was supported by the Ministry of Science and Technology
National Basic Science Program (Project 973) under Grants Nos.
2012CB821804 and 2014CB845806, the Strategic Priority Research
Program "The Emergence of Cosmological Structure" of the Chinese
Academy of Sciences (No. XDB09000000), the National Natural Science
Foundation of China under Grants Nos. 11373014 and 11073005, the
Fundamental Research Funds for the Central Universities and
Scientific Research Foundation of Beijing Normal University, and
China Postdoctoral Science Foundation under grant Nos. 2014M550642
and 2014T70043.

M.B. obtained approval of foreign talent introducing project in
China and gained special fund support of foreign knowledge
introducing project. He also gratefully acknowledges hospitality of
Beijing Normal University.

%M.B. was supported by the Recruitment Program of High-end Foreign Experts in China and he gratefully acknowledges hospitality of
%Beijing Normal University where this research was started.

\end{document}